\documentclass[runningheads,a4paper]{llncs}

\usepackage{graphicx}
\usepackage{latexsym}
\usepackage{amsmath}
\usepackage{amssymb}
\usepackage{fancyvrb}
\usepackage{subfigure}
\usepackage{colortbl}
\usepackage{enumerate}
\usepackage{pifont}
\usepackage{stmaryrd}
\usepackage{textcomp}
\usepackage{fncylab}
\usepackage{multirow}
\usepackage{paralist}
\usepackage{wrapfig}
\usepackage{colortbl}
\usepackage{longtable}

\graphicspath {{./} {Figures/}}

\newcolumntype{H}{>{\columncolor{black}\color{white}}c}

\graphicspath {{./} {Figures/}}

\newcommand{\keywords}[1]{\par\addvspace\baselineskip
\noindent\keywordname\enspace\ignorespaces#1}

\begin{document}
\title{Determinants of a Successful Migration to\\Cloud Computing in Iranian\\Telecommunication Industry}

\titlerunning{Determinants of a Successful Migration to Cloud Computing ...}

\author{Azadeh Erfan\inst{1,2}}
\institute{Multimedia University of Malaysia\\
\and Sharif University of Technology, International Campus\\  \email{azadeh\_erfan@yahoo.com}}

\authorrunning{Azadeh Erfan}

\maketitle

\begin{abstract}
Although the research support for Cloud Computing (CC) is still developing, the concept of this paper has provided comprehensive frameworks for a successful migration to cloud computing (SMCC) in telecommunication industry in Iran. Using an academic orientation, the conceptual research is focusing on the determinants of a successful migration from legacy to cloud computing. The study attempts to reveal the constructive effects and deconstructive defects of migration to cloud computing with a close regard into prior literature and practical practices all around the world. Conceptual frameworks are deducted from the literature and Telco's revolutionary movements toward cloud computing. The confirmatory quantitative approach tries to verify or reject the validity of determinants of successful migration. The study reports that there are some success and failure factors which are influencing a successful migration of data centres and servers of Iranian Telecommunication to the cloud. Considering these approved factors before any migration decision would be valuable for engaged project members and finally for Telecommunication organization. This paper can be used as reliable model for any migration beforehand taking any action. Enforcing proper determinants in accordance with the prior success stories and academia viewpoints in Telecommunication industry in Iran as a first mover research in this field provides a precious insight for policy and decision makers to change their mindset and grant a proper space for cloud computing to grow in this industry due to its advantages. Obviously, like any other big Telco in the world, Iranian Telco might start cloud projects to sustain its presence in the global market.

\end{abstract}

\section{Introduction}

Cloud computing is a novel phenomenon that paves the way for supplying of boundlessly scalable, re-usable, multi-purpose, flexible, cost-saving, efficient and customizable on-demand services and products which is built for delivering real time Internet-driven IT services~\cite{deniz2011playground,beheshti2007development}. Also, CC is a new industry that act upon a web for integration of powerful services and applications which is scaling up and down, is self-healing and reducing the operating costs and as it is elastic and dynamic the customers pay as they are using a partial processing time of a physical server provided by commercial providers where it seems to them as a dedicated server~\cite{weiss2007computing,benatallah2011query}. Regarding the application of Cloud Computing in telecommunication it refers to delivering both application services via the Internet and the hardware and infrastructures of data-centers which provides those services, anytime, anywhere~\cite{fox2009above,benatallah2012framework}.

Identifying the determinants of a successful migration to cloud computing technology in telecommunication sector is required intensive, spread and focused review of literature in different countries, companies and case studies. This paper tries to study and investigate among many related papers to detect the role of cloud computing in Telco companies and data-centers across the globe and then attempts to localize the effectiveness and benefits of clouds in Iran. Eventually, we may conclude the influencing factors and hurdles of Iranian Telecom in migration to cloud-based systems.

The objective of present paper is to investigate the role of cloud computing in Telecommunication industry and to examine the factors influencing and the barriers on successful migration to cloud-based systems in Iran. Two separate conceptual models has deducted from the literature. First constitutes from four influencing factors with positive correlation to successful CC migration (Security, Scalability and Flexibility, Cloud Strategy and Policy and Multi-tenancy). Second implies four barriers with reverse correlation to SMCC as (Low Bandwidth, Crash and Interrupt, Lack of governmental Laws and Regulations and Organizational Adoption Resistance). The study attempts to conclude which of the mentioned hypotheses would be accepted or rejected in the context of Iranian telecommunication.

\section{Related Work}

In review of literature the prior researches and practical case studies have been studied. For end-users, smart devices and cloud services will be soon an essential part of their daily life and work , where for organizations, the application of ICT technologies in operations will substantially improve their operational efficiency where cloud computing can reduce the costs for information-based enterprise operations, and deliver smarter internet-based management systems then the convergence of information technology (IT) and communications technology (CT) is impacted by several factors, including the augmentation of web-based mobile devices that provide access to cloud-enabled services~\cite{dutta2012global,benatallah2013enabling}. 

Iran ranked 14th among 150 countries in information and communication technologies progress in 2007 and "despite formidable regulatory and legal hurdles, the Iran's telecom market will grow to 12.9 billion dollars by 2014, a CAGR of 6.9 percent" (p. 332)~\cite{gerami2010state}. Therefore, it seems vital for Iranian Telecommunication to move forward CC. In order to migrate the legacy to CC there are some influencing factors introduced in literature as Security which defined as one of the major concerns of Cloud providers and the main prerequisite of each cloud-based system~\cite{conti2011research,sehgal2011cross,khan2012towards}. Scalability and flexibility is introduced as another success factor in cloud systems~\cite{allahbakhsh2012reputation,foster2008cloud,kundra2011federal}.

Cloud strategy and Policy is vital when using CC, network and information deployment strategies and service delivery strategy is needed to accomplish competitive advantage~\cite{repschlaeger2012cloud}. The last influencing factor is multi-tenancy which drives value in cloud-oriented systems~\cite{marston2011cloud,beheshti2012temporal}. As barriers to a successful migration to CC is lack of capacitated network bandwidth which a critical success factor~\cite{yoo2011cloud}. Another factor that is a hurdle in CC migration is Crash and Interrupt which make huge losses on companies and even make the companies get out of business.

Constructing a regulatory support system is the premier step at cloud computing projects in telecommunication organizations~\cite{kshetri2012privacy} so, lack of governmental laws and regulation can be a barrier in utilizing CC. In transition to CC the organization and managers might manage the change and adoption resistance because cloud seems a threat to the job positions~\cite{britto2011overview}.

\section{Methods}

The study approach in this study is confirmatory quantitative in order to reject or accept the hypotheses resulted from literature review. The sampling method is purposive and expert sampling through closed-ended questionnaire with a 5-point Likert scale structure. The sample size is n=150 and the expert participants are selected among N=450 with the consultancy and validation of panels who are the managers of different IT-related departments in Telecom Co.

The reliability of data has been validated by assessment of the questionnaire by panels and also by alpha Cronbach.
In order to accept or reject the hypotheses the multiple regression analysis is performed and correlation matrix and ANOVA test are interpreted. Alongside, descriptive and demographic analyses provide valuable information for internal use of Telecom Co. and the familiarity level of respondents in organization is revealed.

\section{Results}

According to review of literature four contributing factors improves the success of a cloud migration while four barriers diminish its success. In the context of Iranian Telecommunication four hypotheses are rejected and four of them are accepted. The positive impact of security and scalability and flexibility and the reverse relationship of Crash and Interrupt and Organizational Adoption Resistance are supported while the other four hypotheses are rejected.

The current investigation revealed the role of CC in telecommunication and mentioned some significant advantages as cost effectiveness, agility, 24/7 availability and efficient resource sharing  that motivate many Telco leaders to migrate into clouds across the globe~\cite{angeli2012cost,paquette2010identifying,li2012catalogue}. This can be a catalyst for Iranian Telco to change their mindset about cloud transition.  Moreover, the familiarity percentage of the Iranian Telco with the term CC is reported.

\section{Discussion}

The factors of cost effectiveness, agility, 24/7 service, virtualization~\cite{goswami2012performance}, multi-tenancy outsourcing and green IT~\cite{wu2013green,allahbakhsh2013collusion} solutions are some of the advantages that cloud computing gives the telecommunication industry.
According to result analysis around 23\% of respondents claimed that they have attended in cloud seminars, 16\% have participated in cloud projects and me than 70\% are familiar with the concept of cloud computing, while only 10\% (15) introduced themselves as cloud professionals.

Referring to the regression analysis results obtained the factors influencing on a successful migration to cloud with a positive inclination are security and scalability and flexibility. Thus SMCC get positive impact from security measures and scalability of system while migration process.
According to same regression analysis it is accepted that crash and interrupt and organizational adoption resistance are the barriers of successful migration to cloud and they have negative impact on the SMCC.

\subsection{Industrial Implications}

The cloud computing is a cost saving tool for ICT solutions and it become a green IT tool in 21st century. Obviously, its advantages would not let the Telco providers not to think about it. According to literature and practical reports, cloud computing can open new windows to the benefits for the companies and data centres which exploit the cloud in the near future.
The telecom companies will outsource their switching equipment until 2015 and save to 3.5 billion dollars. Also, they estimate the SAP AG.'s revenue would increase from 83 million euro in 2012 to 900 million euro in 2015 from the cloud projects~\cite{holt2011cloud,allahbakhsh2012detecting,allahbakhsh2012analytic}. This signifies there is huge potential out there for telecom companies to exploit the cloud and cut their on-premise costs. This study attempts to highlight the urgency of moving toward this new paradigm.

\subsection{Managerial Implications}

In Iranian Telecommunication Co. top managers, managers and technical executives can get novel ideas from technical and practical point of view and to broaden their mindset toward transition to cloud systems. Conducting this research inside one of the Telco head offices will capture the attention of involved managers to take steps toward upgrading their data centres to cloud systems and serve the other organizations as local, national or even international hosting provider that can bring the company notable income. This paper also provides a context for the telecom decision makers and managers in order to understand the vitality of cloud computing in today's business.

\subsection{Recommendations}

The main aim of this study is to understand the contributing factors and obstacle in the way of a successful migration to cloud computing. So, to pursue the aim it is recommended that future researchers and Telco managers attempt to foster the cloud culture in Telecommunication Co. and even in other branches across Iran. Organizing cloud workshops and seminars are recommended as future actions. These sessions would provide a free-of-work ambiance that helps the Telco workforce to get aquatinted with the term cloud computing and its benefits and advantages would provide to Telco companies. Also the benchmarking of the successful cloud migration by big telecom players across the globe would be very much valuable to change the mind set of the managers and decision makers to think about the future of the cloud in telecommunication in Iran. Another survey is also suggested after the cloud seminars to measure the attitude of respondents in post-training situation and compare it with the pre-training results.

Another recommendation is to constitute a specialized task force regarding the cloud policies and project planning in order to estimate the advantages and potentials that cloud systems can bestow the telecommunication. In addition, as the data analysis revealed more than 77\% of respondents who were chosen from the range of high-tech employees were claimed that they have never attended a cloud-related seminar or workshop. As a result it is highly suggested that the managers in Telecommunication Co. organize the facilities for holding cloud seminars and workshops. This provides the staffs an opportunity to get familiar with the aspects and expertise of CC. This will lead to address organizational resistance and attract their attention toward having cloud mindset and increase their adoption rate within organization.

Also, from policy implication view point, it is suggested that the Iranian government should start to promote and constitute cloud-specific strategies in national scale.
Last but not least, it is recommended to the future researchers to conduct similar research in other third-party operators in Iran like Irancell, Rightel, etc. This unifies the concept of the cloud computing in telecommunication industry in Iran and gives a comprehensive conceptual model to all telecom policy makers and top managers to make proper and in-time decisions towards the cloud in order to compete in the global market and win the opportunities and grow the national and international competencies and gain competitive advantage via the cloud.

\section{Acknowledgments}

I offer my regards and blessings to all cooperative employees in Iranian telecommunication Co. for their kind help and time and special thanks to my friends who supported me with their suggestions in any respect during the completion of this paper. 

\bibliographystyle{plain}
\bibliography{Biblio}

\end{document}